\documentstyle[12pt]{article}

\def\be{\begin{equation}}
\def\ee{\end{equation}}

\begin{document}

\setcounter{page}{1}

\rightline{OSU--NT--94--08}
\vskip.5in

\centerline{\large {\bf Light-Front QCD and the Constituent
Quark Model}\footnote{Based on a talk presented by K.G. Wilson at
``Theory of Hadrons and Light-Front QCD,'' Polana Zgorzelisko, Poland,
August 1994.}
}
\vskip.3in
\centerline{ Kenneth G. Wilson and David G. Robertson }
\centerline{\it Department of Physics, The Ohio State University,
Columbus, OH 43210 }

\vskip.6 in
\centerline{\bf Abstract}
\vskip.1in
A general strategy is described for deriving a constituent
approximation to QCD, inspired by the constituent quark model and
based on light-front quantization.  Some technical aspects of the
approach are discussed, including a mechanism for obtaining a
confining potential and ways in which spontaneous chiral symmetry
breaking can be manifested.
\vskip.1in

\vskip.3in
{\bf 1. Introduction}
\vskip.1in

In order to help organize our thinking about QCD and our understanding
of hadronic physics it may be useful to group some relevant issues
into three broad categories. These categories are certainly not meant
to be a complete set, but are meant to help guide us in our approach
to the problems we must face when attempting to solve QCD.

In the first category we have what we might call exact representations
of QCD, for example, the complete set of Schwinger-Dyson equations for
QCD, or the continuum limit of lattice QCD.  We shall also include in
this category ``exact'' light-front theory, by which we mean
light-front quantized QCD including all necessary effects of vacuum
degrees of freedom (also known as ``zero modes,'' though this phrase
has several distinct meanings in light-front quantization). This
theory has a nontrivial vacuum state due to the presence of zero
longitudinal momentum particles.  Correctly incorporating these into
the theory from the beginning is a difficult problem, and is a subject
of ongoing research efforts.

In the second category we have simple pictures of hadronic physics,
each of which may roughly correspond to one or more of the exact
representations.  In this group we have, for example, truncation to
the first Schwinger-Dyson equation, or the strong coupling limit of
lattice QCD.  Corresponding loosely to the exact light-front theory we
have several simple pictures, among them the infinite momentum frame
and the closely related parton model.  The picture we shall
particularly emphasize in connection with light-front physics,
however, is that of the constituent quark model (CQM).  In the CQM
only a minimum number of constituents required by the symmetries are
used to build each hadron, and the vacuum is trivial.  The possibility
of a connection to light-front QCD follows from the simplicity of the
vacuum on the light front, as will be discussed below.  The
strong-coupling limit of lattice QCD is also linked to the constituent
quark model, but in a lattice version.  The CQM discussed here exists
in a continuum, with confining potentials rather than the stringlike
confinement of lattice QCD.

In the third category we have the problem of building bridges between
the simple pictures and the exact representations of QCD.  In lattice
QCD, for example, a bridge between the strong coupling and continuum
limits is provided by the Monte Carlo technique.  The problem we wish
to focus on here is that of building a bridge between the CQM and
exact light-front QCD.

We should emphasize from the outset an important point concerning the
various models and theories.  It is that, while the various exact
representations must be precisely equivalent in the physics they
predict, there is no such requirement for the various simple pictures.
All of these have been obtained by making drastic and sometimes
completely uncontrolled approximations to the full theory, and
afterwards there clearly need be no relation between them. This is not
to say, of course, that understanding in detail the relationship of a
model to the full theory---that is, the bridge---might not lead to
ways of incorporating important physics from others of the models.

Let us begin with the observation that in a certain sense the
discovery of the formal rules of QCD represented a step backwards.  In
the CQM, starting before QCD was developed, hadrons are simple bound
states of a few quark constituents.  The up and down quarks are taken
to have a mass roughly equal to half the average mass of the
nonstrange mesons, around 300 MeV.  Were it present in the model, the
gluon would also be assigned a rather large mass, because we have not
yet seen any low-mass gluonium. The vacuum is trivial.  In this
picture confinement is implemented by an {\it ad hoc} potential.
There is no gluon emission or absorption since there are no gluons.
There is no need to understand the role of renormalization in this
model, because with {\it ad hoc} two-body potentials there is no need
for renormalization.  However, a major failing of the CQM is that it
tells us nothing about mechanisms for chiral symmetry breaking.

QCD was a step backwards in the sense that it forced upon us a complex
and mysterious vacuum.  In QCD, because the effective coupling grows
at long distances, there is always copious production of low-momentum
gluons, which immediately invalidates any picture based on a few
constituents.  Of course, this step was necessary to understand the
nature of confinement and of chiral symmetry breaking, both of which
imply a nontrivial vacuum structure. But for 20 years we have avoided
the question: Why did the CQM work so well that no one saw any need
for a complicated vacuum before QCD came along?

A bridge between equal-time quantized QCD and the equal-time CQM would
clearly be extremely complicated, because in the equal-time formalism
there is no easy {\it nonperturbative} way to make the vacuum simple.
Thus a sensible description of constituent quarks and gluons would be
in terms of quasiparticle states, i.e., complicated collective
excitations above a complicated ground state.  Understanding the
relation between the bare states and the collective states would
involve understanding the full solution to the theory.  On the light
front, however, simply implementing a cutoff on small longitudinal
momenta suffices to make the vacuum completely trivial.  Thus we
immediately obtain a constituent-type picture, in which all partons in
a hadronic state are connected directly to the hadron, instead of
being simply disconnected excitations in a complicated medium.
Whether or not the resulting theory allows reasonable approximations
to hadronic states to be constructed using only a {\it few}
constituents is an open question.  However, we propose to regard the
relative success of the CQM as a reason for optimism.

The price we pay to achieve this constituent framework is that the
renormalization problem becomes considerably more complicated on the
light front.  Counterterms for divergences arising from large
transverse momenta involve entire functions of longitudinal momenta,
and vice versa.  The situation is further complicated by the fact that
the regulators we are forced to use, which must be nonperturbative and
applicable in a Hamiltonian framework, are neither Lorentz nor gauge
invariant.  It is in this way that the familiar ``Law of Conservation
of Difficulty'' manifests itself in our approach.

In the remainder of this paper we shall sketch the approach taken by
the group at OSU, which is presented in more detail in Ref. [1].
(Discussion of some newer developments can be found in the recent
lecture notes of R. Perry [2].)  The bridge that we hope to construct
includes a renormalization procedure that takes us from a field
theoretic bare Hamiltonian to one that is strongly cut off and is much
closer to a CQM-like model.  We shall set up the CQM side of the
bridge by introducing constituent masses for both quarks and gluons,
and then examining the case that the running coupling is small at the
scale of these masses.  We thereby obtain a theory that can be studied
in much the same way bound states are analyzed in QED.  This means
using one perturbation theory to study the renormalization problem and
another to compute bound state properties.  We must in the end
extrapolate back to the physical value of the coupling.  The bridging
to QCD is this extrapolation process.

We shall begin by describing some general aspects of the approach and
the overall strategy.  We then discuss two key ingredients of the
program: first, the appearance of a confinement mechanism starting
from the canonical light-front Hamiltonian for QCD; and second, the
nature and realization of chiral symmetry breaking on the light front.
The problems of confinement and chiral symmetry breaking have been the
two major barriers to building the bridge we seek, because they
prevented us from making a straightforward connection to a
weak-coupling picture.  We hope to explain how these barriers may be
overcome in light-front QCD.  There are many other important details
and caveats that we shall not discuss here; we present only some bare
essentials in oversimplified form.  The reader is referred to Ref. [1]
for more details.

\vskip.3in
{\bf 2. Generalities}
\vskip.1in

In order to establish some notation and to have a starting point for
the discussion, let us begin by sketching the canonical light-front
Hamiltonian for QCD.  We choose the light-cone gauge $A^+_a=0$ and
finesse (i.e., ignore for now) the problem of the zero modes.

There is first of all a free part
\be
H_{free}= \int dk^+ d^2k_\perp \biggl\{
\sum_\lambda { k_\perp^2+\mu^2 \over k^+} a^\dagger_{k\lambda}
a_{k\lambda}
+ \sum_\sigma { k_\perp^2+m^2 \over k^+}
\Bigl(b^\dagger_{k\sigma} b_{k\sigma}
+d^\dagger_{k\sigma} d_{k\sigma} \Bigr) \biggr\} \; ,
\ee
which counts the light-front energy of each constituent.  Here
$\lambda$ and $\sigma$ index gluon and quark polarizations,
respectively, and color indices have been suppressed.  Note that we
have included a mass term for the gluons as well as the quarks
(although we include only transverse polarization states for the
gluons).  We have in mind here that all masses that occur in
$H_{free}$ should roughly correspond to constituent rather than
current masses.  There are two points that should be emphasized in
this regard.

First, cutoff-dependent masses for both the quarks and gluons will be
needed anyway as counterterms.  This occurs because all the cutoffs we
have at our disposal for nonperturbative Hamiltonian calculations
violate both equal-time chiral symmetry and gauge invariance.  These
symmetries, if present, would have protected the quarks and gluons
from acquiring this kind of mass correction.  Instead, in the
calculations we are discussing both the fermion and gluon self-masses
are quadratically divergent in a transverse momentum cutoff $\Lambda$.

The second point is more physical. When setting up perturbation theory
(more on this below) one should always keep the zeroth order problem
as close to the observed physics as possible.  Furthermore, the
division of a Hamiltonian into free and interacting parts is always
completely arbitrary, though the convergence of the perturbative
expansion may hinge crucially on how this division is made.  Nonzero
constituent masses for both quarks and gluons clearly comes closer to
the phenomenological reality (for hadrons) than do massless gluons and
nearly massless light quarks.

Next we have the various interactions, among which are the standard
quark-gluon and gluon self-couplings.  There are also
``instantaneous'' interactions, four-point operators that arise when
certain constrained field components are eliminated by their equations
of motion.  Elimination of $A^-_a$ results in the instantaneous gluon
interactions:
\begin{eqnarray}
H_{g} &=& - 2g^2 \int dx^- d^2 x_\perp
\biggl\{ (\psi_+^{\dagger} T^a \psi_+ ) \left( \frac{1}
{\partial^+} \right)^2 (\psi_+^{\dagger} T^a \psi_+) \\
\nonumber\\
 & & +(f^{abc} A_b^i \partial^+ A_c^i)
\left( \frac{1}{\partial^+} \right)^2 (\psi_+^{\dagger} T^a\psi_+)
+\frac{1}{4} f^{abc} f^{ade}
 (A_b^i \partial^+ A_c^i) \left(
		\frac{1}{\partial^+} \right)^2 (A_d^j \partial^+ A_e^j)
		\biggl\}\; , \nonumber
\end{eqnarray}
while elimination of the constrained components of $\psi$ gives rise
to the instantaneous fermion term:
\be
H_{f} = g^2 \int dx^-d^2x_\perp \thinspace
\psi_+^{\dagger} \alpha_\perp \cdot
          A_{\bot} \left( \frac{1}{i \partial^+}
           \right) \alpha_\perp \cdot
            A_{\bot} \psi_+ \; .
\ee
Here $\psi_+\equiv\frac{1}{2}\gamma^0\gamma^+\psi$ is the dynamical
part of the Fermi field, and the $T^a$ are the generators and the
$f^{abc}$ the structure constants of color $SU(3).$

Now, the presence of a nonzero gluon mass has important consequences.
First, it automatically stops the running of the coupling below a
scale comparable to the mass itself.  This allows us to (arbitrarily)
start from a small coupling at the gluon mass scale so that
perturbation theory is everywhere valid, and only extrapolate back to
the physical value of the coupling at the end.  The quark and gluon
masses also provide a kinematic barrier to parton production; the
minimum free energy that a massive parton can carry is ${m^2\over
p^+}$, so that as more partons are added to a state and the typical
$p^+$ of each parton becomes small, the added partons are forced to
have high energies.  Finally, the gluon mass eliminates any infrared
problems of the conventional equal-time type.

Next let us discuss cutoffs.  When contemplating cutoffs there are
certain general issues to be considered, for example whether to mix
the transverse and longitudinal momentum cutoffs or keep them
separate.  This question is sharpened with the realization that
longitudinal scale invariance must be an exact Lorentz symmetry of the
theory, while transverse scale invariance is broken by masses and the
cutoff itself.  For the present we shall imagine a simple cutoff on
constituent energies, that is, requiring
\begin{equation}
{p_\perp^2+m^2\over p^+} < {\Lambda^2\over P^+}
					\label{eq:cutoff}
\end{equation}
for each constituent in a given Fock state.  Here $P^+$ is a parameter
that sets a longitudinal scale.\footnote{We write the cutoff in this
way in order to make explicit that the RHS of (\ref{eq:cutoff}), which
in normal rest frames is measured in units of mass, actually scales
like a transverse mass squared divided by a longitudinal mass.  For a
discussion of the scaling behavior of various quantities and power
counting on the light front see Ref. [1].} Note that some Fock states
have high energies due to the presence of a few high-energy partons,
while others have high energies due to the presence of very many
low-energy partons.  Because of the presence of masses, however, Eq.
(\ref{eq:cutoff}) results in a small-$p^+$ cutoff, so that the total
number of partons in a state of given total $p^+$ is bounded.  Thus
the full quantum field theory is reduced to an ordinary quantum
mechanical many-body problem.

Imposing (\ref{eq:cutoff}) does not completely regulate the theory,
however; there are additional small-$p^+$ divergences coming from the
instantaneous terms in the Hamiltonian.  We shall regulate these by
treating them as if the instantaneous exchanged gluons and quarks were
actually constituents, and were required to satisfy condition
(\ref{eq:cutoff}).

Other cutoff schemes are of course also possible, and may in fact be
preferable for more refined calculations than the ones we shall
discuss here. One can impose a cutoff on the invariant mass of Fock
states, which is potentially useful because it respects those Lorentz
symmetries that are kinematic on the light front, in particular
longitudinal boost invariance.  A disadvantage of this scheme is that
it leads to spectator-dependence in counterterms.  Another possibility
is based on a transverse lattice, perhaps coupled with a momentum
discretization in the $(x^+,x^-)$ plane.

We can now outline the first stage of the program.  Having stopped the
running of the coupling below the constituent mass scale, we
arbitrarily take it to be small at this scale, so that perturbation
theory is valid at all energy scales.  We can now use power counting
to identify all relevant and marginal operators (relevant or marginal
in the renormalization group sense).  Because of the cutoffs we must
use, these operators are not restricted by Lorentz or gauge
invariance.  Because we have forced the vacuum to be trivial, the
effects of spontaneous chiral symmetry breaking must be manifested in
explicit chiral symmetry breaking effective interactions.  (We shall
return to this below.) This means the operators are not restricted by
chiral invariance either.  There are thus a large number of allowed
operators.  Furthermore, since transverse divergences occur for any
longitudinal momentum, the operators that remove transverse cutoff
dependence contain functions of dimensionless ratios of all available
longitudinal momenta.  That is, many counterterms are not
parameterized by single coupling constants, but rather by entire
functions of longitudinal momenta.  A precisely analogous result
obtains for the counterterms for light-front infrared divergences;
these will involve entire functions of transverse momenta.

The counterterm functions can in principle be determined by requiring
that Lorentz and gauge invariance be restored in the full theory.
Alternatively, we might invoke the idea of coupling coherence [3].
The basic idea is that the coefficients of the operators that appear
in the effective Hamiltonian are not arbitrary; they are in principle
computable in terms of the one coupling parameter $g$ that
characterizes the relativistic theory, along with the quark mass
parameters. Thus as the cutoff is varied the running of these
coefficients should occur only through their dependence on a single
running coupling constant $g_\Lambda$.  Requiring this to be the case
can be used to fix many of the couplings.  In fact, in all examples
worked out so far coupling coherence automatically yields the correct
counterterms necessary to restore, e.g., Lorentz invariance, and in
the case of the $\sigma$ model can also be used to fix the strengths
of symmetry breaking operators that arise as a result of spontaneous
symmetry breaking.

The cutoff Hamiltonian, with renormalization counterterms, will thus
be given as a power series in $g_\Lambda$:
\begin{equation}
H(\Lambda) = H^{(0)} + g_\Lambda H^{(1)}+g_\Lambda^2 H^{(2)}
+ \dots\; ,
\end{equation}
where all dependence on the cutoff $\Lambda$ occurs through the
running coupling $g_\Lambda,$ and cutoff-dependent masses.  The next
step is to make this Hamiltonian look more like that of a CQM by
performing a similarity transformation on it.

\vskip.3in
{\bf 3. A CQM-Like Low-Energy Hamiltonian }
\vskip.1in

The next stage in building a bridge from the CQM to QCD is to
establish a connection between the {\it ad hoc} $q\overline{q}$
potentials of the CQM and the complex many-body Hamiltonian of QCD
(see Eqs. (2), (3), and (5)).  We shall illustrate in a very
simplified way how this connection can be established.  The full
formalism needed is described in Ref. [4].

In lowest order the canonical QCD Hamiltonian contains gluon emission
and absorption terms, including emission and absorption of high-energy
gluons.  Since a gluon's energy is ${k_\perp^2+\mu^2 \over k^+}$ for
momentum $k$, a high-energy gluon can result either if $k_\perp$ is
large or $k^+$ is small.  But in the CQM, gluon emission is ignored
and only low-energy states matter.  How can one overcome this double
disparity?  The answer is that we can change the initial cutoff
Hamiltonian $H(\Lambda)$ by applying a unitary transformation to it.
We imagine constructing a transformation $U$ that generates a new
effective Hamiltonian $H_{\rm eff}$:
\be
H_{\rm eff} = U^\dagger H(\Lambda) U\; .
					\label{eq:simtrans}
\ee
We then {\it choose} $U$ to cause $H_{\rm eff}$ to look as much like a
CQM as we can.

The essential idea is to start out as though we were going to
diagonalize the Hamiltonian $H(\Lambda),$ except that we stop short of
computing actual bound states.  A complete diagonalization would
generate an effective Hamiltonian $H_{\rm eff}$ in diagonal form; all
its off-diagonal matrix elements would be zero.  Furthermore, in the
presence of bound states the fully diagonalized Hamiltonian would act
in a Hilbert space with discrete bound states as well as continuum
quark-gluon states.  In a confined theory there would only be bound
states.  What we seek is a compromise: an effective Hamiltonian in
which some of the off-diagonal elements can be nonzero, but in return
the Hilbert space for $H_{\rm eff}$ remains the quark-gluon continuum
that is the basis for $H(\Lambda)$.  No bound states should arise. All
bound states are to occur through the diagonalization of $H_{\rm
eff}$, rather than being part of the basis in which $H_{\rm eff}$
acts.

To obtain a CQM-like effective Hamiltonian, we would ideally eliminate
all off-diagonal elements that involve emission and absorption of
gluons or of $q\overline{q}$ pairs.  It is the emission and absorption
processes that are absent from the CQM, so we should remove them by
the unitary transformation.  However, we would allow off-diagonal
terms to remain within any given Fock sector, such as $q\overline{q}
\rightarrow q\overline{q}$ off-diagonal terms or
$qqq \rightarrow qqq$ terms.  This means we allow off-diagonal
potentials to remain, and trust that bound states appear only when the
potentials are diagonalized.

Actually, as discussed in Ref. [4], we cannot remove all the
off-diagonal emission and absorption terms.  This is because the
transformation $U$ is sufficiently complex that we only know how to
compute it in perturbation theory.  Thus we can reliably remove in
this way only matrix elements that connect states with a large energy
difference; perturbation theory breaks down if we try to remove, for
example, the coupling of low-energy quark to a low-energy quark-gluon
pair.  We therefore introduce a second cutoff parameter
${\lambda^2\over P^+}$, and design the similarity transformation to
remove off-diagonal matrix elements between sectors where the energy
{\it difference} between the initial and final states is greater than
this cutoff.  For example, in second order the effective Hamiltonian
has a one-gluon exchange contribution in which the intermediate gluon
state has an energy above the running cutoff.  Since the gluon energy
is ${k_\perp^2+\mu^2\over k^+}$, where $k$ is the exchanged gluon
momentum, the cutoff requirement is
\be
{k_\perp^2+\mu^2\over k^+}> {\lambda^2\over P^+}\; .
\ee
This procedure is known as the ``similarity renormalization group''
method.  For a more detailed discussion and for connections to
renormalization group concepts see Ref. [4].

\vskip.3in
{\bf 4. Solving $H_{\rm eff}$: Two Perturbation Theories }
\vskip.1in

The result of the similarity transformation is to generate an
effective light-front Hamiltonian $H_{\rm eff},$ which must be solved
nonperturbatively.  Guided by the assumption that a constituent
picture emerges, in which the physics is dominated by potentials in
the various Fock space sectors, we can proceed as follows.

We first split $H_{\rm eff}$ anew into an unperturbed part $H_0$ and a
perturbation $V$.  The principle guiding this new division is that
$H_0$ should contain the most physically relevant operators, e.g.,
constituent-scale masses and the potentials that are most important
for determining the bound state structure.  All operators that change
particle number should be put into $V$, as we anticipate that
transitions between sectors should be a small effect.  This is
consistent with our expectation that a constituent picture results,
but this must be verified by explicit calculations.  Next we solve
$H_0$ nonperturbatively in the various Fock space sectors, using
techniques from many-body physics.  Finally, we use bound-state
perturbation theory to compute corrections due to $V$.

We thus introduce a second perturbation theory as part of building the
bridge.  The first perturbation theory is that used in the computation
of the unitary transformation $U$ for the incomplete diagonalization.
The second perturbation theory is used in the diagonalization of
$H_{\rm eff}$ to yield bound-state properties.  R. Perry in particular
has emphasized the importance of distinguishing these two different
perturbative treatments [2].  The first is a normal field-theoretic
perturbation theory based on an unperturbed free field theory.  In the
second perturbation theory a different unperturbed Hamiltonian is
chosen, one that includes the dominant potentials that establish the
bound state structure of the theory.  Our working assumption is that
the dominant potentials come from the lowest-order potential terms
generated in the perturbation expansion for $H_{\rm eff}$ itself.
Higher-order terms in $H_{\rm eff}$ would be treated as perturbations
relative to these dominant potentials.

It is only in the second perturbative analysis that constituent masses
are employed for the free quark and gluon masses.  In the first
perturbation theory, where we remove transitions to high-mass
intermediate states, it is assumed that the expected field theoretic
masses can be used, i.e., near-zero up and down quark masses and a
gluon mass of zero.  Because of renormalization effects, however,
there are divergent mass counterterms in second order in $H(\Lambda).$
$H_{\rm eff}$ also has second-order mass terms, but they must be
finite---all divergent renormalizations are accomplished through the
transformation $U$.  When we split $H_{\rm eff}$ into $H_0$ and $V$,
we include in $H_0$ both constituent quark and gluon masses and the
dominant potential terms necessary to give a reasonable qualitative
description of hadronic bound states.  Whatever is left in $H_{\rm
eff}$ after subtracting $H_0$ is defined to be $V$.

In both perturbation computations the same expansion parameter is
used, namely the coupling constant $g$.  In the second perturbation
theory the running value of $g$ measured at the hadronic mass scale is
used.  In relativistic field theory $g$ at the hadronic scale has a
fixed value $g_s$ of order one; but in the computations an expansion
for arbitrarily small $g$ is used.  It is important to realize that
covariance and gauge invariance are violated when $g$ differs from
$g_s$; the QCD coupling at any given scale is not a free parameter.
These symmetries can only be fully restored when the coupling at the
hadronic scale takes its physical value $g_s$.

To assure maximum effectiveness of the perturbative computations it is
convenient to alter the coupling dependencies of major terms in the
interaction $V.$ For example, $V$ contains the difference between the
constituent quark and gluon masses used in $H_0$ and the masses in
$H_{\rm eff}$ itself, a difference which is of order one.  But since
this difference only has to be exact for $g=g_s$ we can convert this
term to a second order term by multiplying it by $(g/g_s)^2,$ turning
it into a genuine perturbation.  Likewise some of the second-order
potential terms in $H_0$ (second order meaning ${\cal O}(g^2)$) can be
multiplied by $(g/g_s)^2$ to make them of fourth order in $g$ when $g$
is small, if this is necessary to simplify the bound state analysis
for small $g.$ These modifications can also cause violations of
covariance for $g\neq g_s.$

\vskip.3in
{\bf 5. Confinement }
\vskip.1in

The conventional wisdom is that any weak-coupling Hamiltonian derived
from QCD will have only Coulomb-like potentials, and certainly will
not contain confining potentials.  Only a strong-coupling theory can
exhibit confinement.

The conventional wisdom is wrong.  Robert Perry made this discovery,
as he will report in his contribution to these proceedings [2,5].
When $H_{\rm eff}$ is constructed by the unitary transformation of Eq.
(\ref{eq:simtrans}), with $U$ determined by the ``similarity
renormalization group'' method, $H_{\rm eff}$ has an explicit
confining potential already in second order!  We shall explain this
result below.  However, first we should give the bad news.  If quantum
electrodynamics (QED) is solved by the same process as we propose for
QCD, then the effective Hamiltonian for QED has a confining potential
too.  In the electrodynamic case, the confining potential is purely an
artifact of the construction of $H_{\rm eff}$, an artifact which
disappears when the bound states of $H_{\rm eff}$ are computed.  Thus
the key issues, discussed below, are to understand how the confining
potential is cancelled in the case of electrodynamics, and then to
establish what circumstances would prevent a similar cancellation in
QCD.

The confining potential that appears in $H_{\rm eff}$ in both QED and
QCD is easily understood if we recall a cancellation that occurs
between the infrared divergences of the instantaneous gluon (or
photon) exchange potential and the perturbative one gluon exchange
diagram. (This same cancellation occurs for photons in QED.)  Stripped
to its essence, the terms that cancel behave like
\begin{eqnarray}
{\rm Instantaneous~exchange:} &\quad &-{1\over (q^+)^2} \\
{\rm One~gluon~exchange:} &  &{1\over (q^+)^2}
{q_\perp^2\over q_\perp^2} + {\rm terms~of~order~}{1\over q^+}
\end{eqnarray}
where $q^+$ is the exchanged longitudinal momentum and $q_\perp$ is
the exchanged transverse momentum.  The ratio $(q_\perp^2/q_\perp^2)$
results only for small $q^+$; otherwise the denominator is more
complicated.  When the two terms are added the $({1\over q^+})^2$
terms cancel, leaving no terms more singular than ${1\over q^+}$ at
small $q^+.$

In the similarity renormalization group procedure, the explicit one
gluon exchange diagram is generated only for $q^+$ and $q_\perp$
values for which the gluon energy is greater than the running energy
cutoff ${\lambda^2\over P^+}.$ The gluon energy is ${q_\perp^2\over
|q^+|},$ where the absolute value of $q^+$ appears to keep the energy
positive and no mass occurs because the gluon is massless in the first
stage computation of $H_{\rm eff}$.  As a result of this energy
formula, the instantaneous gluon exchange term $({1\over q^+})^2$
remains uncancelled if ${q_\perp^2\over|q^+|}$ is less that
${\lambda^2\over P^+}.$ This corresponds to a potential in position
space that behaves like
\be
v(x^-,x_\perp) = \int_{-\epsilon}^\epsilon dq^+
\int_0^{|q_\perp|<\sqrt{\lambda^2|q^+|/P^+}} d^2q_\perp
e^{iq^+x^--iq_\perp\cdot x_\perp} {1\over (q^+)^2}\; .
\ee
(We have arbitrarily bounded the $q^+$ integration between
$\pm\epsilon$ because only the small-$q^+$ behavior of the potential
has been computed.  The remainder of the potential will not be
important.)  This potential has a divergent constant term: when
$x^-=x_\perp=0,$
\be
v(0,0) = \int_{-\epsilon}^\epsilon dq^+
{\pi \lambda^2|q^+|\over P^+}{1\over (q^+)^2}\; ,
\ee
which is logarithmically divergent.  It turns out that this divergence
cancels against self-energy divergences for color-singlet states, and
is therefore irrelevant.  We then compute the difference
$v(x^-,x_\perp) - v(0,0).$ The $q_\perp$ integration when carried out
gives
\be
v(x^-,x_\perp)-v(0,0) = \int_{-\epsilon}^\epsilon dq^+
{\pi \lambda^2|q^+|\over P^+}\biggl\{ e^{iq^+x^-}
f\Bigl(\sqrt{{\lambda^2|q^+|\over P^+}}|x_\perp|\Bigr) - 1\biggr\}
{1\over(q^+)^2}\; ,
\ee
where
\be
f(|y_\perp|) \equiv {1\over\pi}\int_0^{|q_\perp|<1} d^2q_\perp
e^{iq_\perp\cdot y_\perp}\; .
\ee
If either $x^-$ or $|x_\perp|$ is very large (the appropriate limit to
study for confinement) then one can verify that the exponential term
\be
e^{iq^+x^-}
f\Bigl(\sqrt{{\lambda^2|q^+|\over P^+}}|x_\perp|\Bigr)
\ee
is approximately unity if $q^+$ is so small that both $|q^+x^-|$ and
$\sqrt{{\lambda^2|q^+|\over P^+}}|x_\perp|$ are small, so that in this
limit the integrand is negligible.  However, once $q^+$ is large
enough so that either $|q^+x^-|$ or $\sqrt{{\lambda^2|q^+|\over
P^+}}|x_\perp|$ is large, then the exponential term is negligible and
we are left with a logarithmic integral:
\be
-\int_{-\epsilon}^\epsilon
dq^+ {\pi\lambda^2\over P^+} {1\over|q^+|}\qquad
\biggl( |q^+|>{1\over|x^-|}\; ,\; |q^+|>{P^+\over\lambda^2|x_\perp|^2}
\biggr)\; .
\ee
The result is that the potential grows logarithmically whenever either
$|x^-|$ or $|x_\perp|$ or both are large!  If
${\lambda^2|x_\perp|^2\over P^+} > |x^-|$ then the potential behaves
as $\ln({\lambda^2|x_\perp|^2\over P^+})$, otherwise it behaves as
$\ln|x^-|.$ Thus we obtain a logarithmic confining potential along any
direction in the three-dimensional $(x^-,x_\perp)$ space.  For further
details and discussion see Refs. [2,5].

Now we come to the crucial question.  How is this confining potential
cancelled in the case of QED, and why won't the same cancellation
mechanism apply to QCD also?

In the case of QED the cancellation comes from the obvious source:
one-photon intermediate states with photon energies below the cutoff.
The effective Hamiltonian $H_{\rm eff}$ for QED contains one photon
emission and absorption matrix elements for photons with energies
below the cutoff.  When the remaining one-photon states are eliminated
by perturbation theory, the confining potential is cancelled and the
normal Coulomb potential is all that remains.

What about QCD?  The corresponding states would be $q\overline{q}g$
states (where $q$ stands for a quark and $g$ a gluon) with gluon
energies below the cutoff ${\lambda^2\over P^+}.$ Now we must be
careful, however.  If we imagine eliminating such states by
perturbation theory we will obtain a contribution of the usual
second-order form
\be
{ \langle q\overline{q}|V|q\overline{q}g\rangle
\langle q\overline{q}g|V|q\overline{q}\rangle\over
E_{q\overline{q}} - E_{q\overline{q}g}}\; ,
\ee
where $E$ is the energy of an unperturbed state.  This assumes that we
start with a $q\overline{q}$ state and make a transition to a
$q\overline{q}g$ state.  The most simple-minded computation we can
make with this perturbative form is to use only the free part of $H_0$
to define both the $q\overline{q}$ and $q\overline{q}g$ states and
their unperturbed energies $E_{q\overline{q}}$ and
$E_{q\overline{q}g}$.  In this case the energy denominator reduces to
the gluon energy.  But now our expectation for QCD, if it represents
the physics of hadrons correctly, is that only a rather large gluon
mass can come close to representing the physics.  Hence the energy
denominator would include a factor $q_\perp^2+\mu^2,$ where $\mu$ is
the gluon mass.  The presence of this mass changes the one-gluon
exchange term from
\be
{1\over(q^+)^2} {q_\perp^2\over q_\perp^2}
\ee
to
\be
{1\over(q^+)^2} {q_\perp^2\over q_\perp^2+\mu^2}\; .
\ee
This means that if $|q_\perp|$ is much less than $\mu$ the one-gluon
exchange term determined from $H_{\rm eff}$ is far too small to cancel
the instantaneous term.  However, when one examines the range of
$q_\perp$ that contributes significantly to the logarithmic growth of
the potential as $x^-$ or $x_\perp$ become large, one finds that this
range involves only very small values of $|q_\perp|$, small enough so
that one-gluon exchange is negligible.

This argument is very simplistic, based on an artificially constructed
free Hamiltonian with all potentials ignored.  Examining the analysis
with care, the crucial question is to determine the difference
$E_{q\overline{q}} - E_{q\overline{q}g}$ when the $q\overline{q}g$
state includes an infrared gluon: a gluon with $q^+$ and $q_\perp$
small, and with ${q_\perp^2\over q^+}$ of order the cutoff
${\lambda^2\over P^+}.$ This means looking at the full effective
Hamiltonian $H_{\rm eff}$ in the $q\overline{q}$ sector (including
potential terms) to determine $q\overline{q}$ eigenvalues and then
examining the $q\overline{q}g$ sector, including potentials also, to
determine how much the presence of an infrared gluon increases the
energy eigenvalues obtained.  Is this increase enough to prevent
cancellation of instantaneous gluon exchange when $q^+$ and $q_\perp$
are both small?  Our working assumption is that at the relativistic
value of the coupling $g_s$, the $q\overline{q}g$ sector will indeed
have sufficiently higher energy eigenvalues than the $q\overline{q}$
sector to block cancellation.

\vskip.3in
{\bf 6. Spontaneous Chiral Symmetry Breaking }
\vskip.1in

Another area of crucial importance to the CQM-QCD bridge is the issue
of spontaneous chiral symmetry breaking. There are two troubling
problems regarding spontaneous chiral symmetry breaking.  The first is
that the vacuum is often presumed to be trivial in light-front field
theory.  But how can a trivial vacuum be also a spontaneously broken
vacuum of any symmetry?  The second question is how any spontaneous
breaking effects could appear in a weak-coupling approach, given that
in the weak coupling limit there is no basis for spontaneous breaking
of $SU(3)\times SU(3)$ to occur?

There is a third question which is troubling to many newcomers to
light-front physics, namely how can one talk about chiral symmetry at
all in an effective Hamiltonian which has nonzero constituent masses
for the up and down quarks even when chiral symmetry is supposed to be
exactly conserved?  This last question puzzles only newcomers to the
field because veterans of light-front research know that on the light
front chiral symmetry is the same as helicity conservation, and in the
free-field limit this is an exact symmetry for any value of the quark
mass.  The only breaking of light-front chiral symmetry occurs in one
of the gluon emission and absorption terms, namely the term that is
linear in the quark mass and explicitly combines helicity flip with
gluon emission or absorption.  Thus the question of chiral symmetry
breaking in $H_{\rm eff}$ concerns the interactions in $H_{\rm eff}$,
not its free field structure.

One often turns to the $\sigma$ model for inspiration on issues of
spontaneous chiral symmetry breaking.  In this model the mechanism
behind the spontaneous breaking is that the $\sigma$ field acquires a
vacuum expectation value (VEV).  In momentum space it is the zero
momentum mode of $\sigma$ that acquires the VEV.  However, all claims
that the vacuum of a light front field theory is trivial depend on
elimination or disregard of the zero momentum modes.  If these modes
are allowed, an arbitrarily complex vacuum built of zero longitudinal
momentum constituents becomes possible.  In the case of the $\sigma$
model one either ($a$) includes the zero momentum modes, making it
possible to have a nontrivial vacuum in which $\sigma$ has a VEV, or
($b$) declares the zero modes to be absent but permits explicit chiral
symmetry breaking terms to appear in the Hamiltonian, of the kind a
shift of $\sigma$ by a constant would generate.  See Appendix A of
Ref. [1] for more details.

By analogy with the $\sigma$ model we propose that zero momentum modes
have an equally crucial role in spontaneous chiral symmetry breaking
for QCD.  After some extensive search we have established a tentative
working assumption, that spontaneous chiral symmetry breaking is
represented by a shift in the zero longitudinal momentum mode of the
{\it lower} (constrained) components of $\psi$ and $\overline{\psi}$.
These shifts are not constants, however.  Instead they are shifts by
composite operators of the form $\psi A$ or $\overline{\psi} A$, with
appropriate spin matrices.  These shifts lead to new explicit chiral
breaking interactions in the canonical QCD Hamiltonian, in particular
new chiral-breaking components in the instantaneous quark exchange
term in the canonical Hamiltonian.

It is only a working assumption that spontaneous chiral symmetry
breaking is represented by explicit chiral breaking terms in the
instantaneous quark exchange term.  The proof has to be that there
exists a locally conserved chiral current despite the explicit
breaking terms.  Unfortunately, we can only expect exact current
conservation when the coupling constant has its relativistic value
$g_s$.  For all smaller values of $g$ there will be explicit
nonconservation of the currents.  Thus until we can extrapolate
reliably from small $g$ to $g_s$ we will have no way of testing our
working assumptions about how spontaneous chiral symmetry breaking is
realized.

\vskip.3in
{\bf 7. Concluding Remarks}
\vskip.1in

The construction of the QCD effective Hamiltonian described above has
already been carried out through second order in perturbation theory
[1].  The resulting theory is closely analogous to the
phenomenological CQM.  The next important step to be taken is to
extend this work to fourth order in $g$, where we expect to begin to
see the phenomenology replaced by true QCD effects.  This step is also
important to see whether the program is practical and whether the
results actually improve.  A useful first test of these Hamiltonians
might be in the study heavy quark systems, where one presumably does
not have to have complete control over confinement and chiral symmetry
breaking to obtain reasonable results.

There are of course other important technical aspects of the
light-front approach that we have not touched on here.  The
renormalization issues in particular have only been sketched very
superficially.  There has been significant work on formulating other
light-front renormalization groups in addition to the similarity
renormalization scheme [6].  Power counting on the light front is
significantly different than at equal times, and is discussed in Ref.
[1].

There are many open problems, both of a practical and conceptual
nature.  One area that has not received sufficient attention concerns
the nature of instantons and the resolution of the $U(1)_A$ problem in
the light front framework.  There are also the problems associated
with the vacuum degrees of freedom, or zero modes.  In the formulation
we have described these are excluded from the outset, so there can be
no direct confrontation of dynamical effects produced by these degrees
of freedom.  Work on incorporating them into the theory should help us
understand the kinds of operators that can appear in the effective
Hamiltonian for the cutoff theory.  Finally, work on constructing
nonperturbative realizations of the renormalization group on the light
front has barely begun.

\vskip.3in
{\bf Acknowledgements}
\vskip.1in

The ideas presented here are the fruit of four years of struggle and
effort by the light-front group at OSU.  A visit by Stan G{\l}azek was
crucial in forcing the whole group to confront the difficult issues
addressed in my talk.  I am especially grateful to Stan G{\l}azek,
Robert Perry, and Avaroth Harindranath for their sustained efforts to
make sense of all the complexities.  I thank everyone else at OSU for
their help---Steve Pinsky, Junko Shigemitsu, Tim Walhout, Wei-Min
Zhang, and Daniel Mustaki, as well as Peter Lepage.

\vskip.3in
{\bf References}
\vskip.1in

1. K.G. Wilson, {\it et al.}, Phys. Rev. D{\bf 49}, 6720 (1994).

2. R.J. Perry, Lectures presented at {\it Hadrons 94,} Gramado,
Brazil, April, 1994.

3. R.J. Perry and K.G. Wilson, Nucl. Phys. B{\bf 403}, 587 (1993).

4. St.D. G{\l}azek and K.G. Wilson, Phys. Rev. D{\bf 48}, 5863 (1993);
{\it ibid. \bf 49}, 4214\par\quad (1994).

5. R.J. Perry, these proceedings.

6. R.J. Perry, Ann. Phys. {\bf 232}, 116 (1994).

\end{document}